\documentclass{article}
\usepackage{spconf,amsmath,graphicx,algpseudocode,multirow,makecell,colortbl}
\usepackage{xcolor}

\title{LPC Augment: An LPC-based ASR Data Augmentation Algorithm for Low and Zero-Resource Children's Dialects}
\name{Alexander Johnson$^1$, Ruchao Fan$^1$, Robin Morris$^2$, and, Abeer Alwan$^1$}
\address{$^1$University of California, Los Angeles, Department of Electrical and Computer Engineering \\ $^2$Georgia State University, Department of Psychology}
\begin{document}
\ninept
\maketitle
\begin{abstract}
This paper proposes a novel linear prediction coding-based data augmentation method for children's low and zero resource dialect ASR.  The data augmentation procedure consists of perturbing the formant peaks of the LPC spectrum during LPC analysis and reconstruction.  The method is evaluated on two novel children's speech datasets with one containing California English from the Southern California Area and the other containing a mix of Southern American English and African American English from the Atlanta, Georgia area.  We test the proposed method in training both an HMM-DNN system and an end-to-end system to show model-robustness and demonstrate that the algorithm improves ASR performance, especially for zero resource dialect children's task, as compared to common data augmentation methods such as VTLP, Speed Perturbation, and SpecAugment.
\end{abstract}
\section{Introduction}
\label{sec:intro}

Current state-of-the-art automatic speech recognition (ASR) systems are unreliable for low-resource dialects \cite{s1}.  Recent approaches have made strides in domain adversarial learning \cite{a1}, cross-dialect transfer learning \cite{a2}, and text-to-speech-based  data augmentation \cite{a3} for dialect and accent-robust ASR. While these approaches are promising, it is also advantageous to create less computationally expensive data augmentation methods that can still make the training samples representative of real-use scenarios.
Data augmentation methods like vocal tract length perturbation (VTLP) \cite{b8}, speed perturbation \cite{b9}, and SpecAugment \cite{b11} seek to create artificial training data containing deviations from the original samples to make the model less sensitive to expected variations. However, these current state-of-the-art data augmentation methods do not specifically target formant shifts and pronunciation differences that are known to occur in some dialects.  In order to make ASR systems more robust to dialectal differences, new data augmentation methods are needed.

In this paper, we attempt to improve children's ASR systems for a low and zero resource dialect.  Children's ASR technology has a large number of uses in early education ranging from literacy practice to pronunciation training \cite{b12}.  However, despite their benefits, ASR systems for children have developed much less quickly and give worse performance than ASR systems for adults due to a number of challenges:  First, children possess less fine motor control over their speech articulators, leading to higher intra-speaker variability \cite{b13,b14}.  Second, as children grow, their vocal tract sizes change at different rates, leading to higher inter-speaker variability \cite{b15,b16}.  Third, procuring large amounts of labeled children's speech data is often difficult. Hence, there is typically not enough children's data to perform robust data-driven ASR experiments.  Even the available children's speech corpora, such as the OGI kids' corpus \cite{b17}, only contain kids' speech of a single dialect, making it difficult to utilize them in training a children's ASR system for less standard dialects like Southern American English or African American English (AAE).  As end-to-end ASR systems with no explicit pronunciation models become widespread, it becomes important to develop more model-invariant data augmentation techniques that can incorporate linguistic knowledge into the model training. Here, we propose a novel data augmentation strategy for improving the performance of both hybrid and end-to-end systems for the understudied dialects of Southern American English and AAE in children given little to no in-domain data.

Data augmentation methods have been largely successful in improving children's ASR performance \cite{b19,b20}. This paper proposes a new linear prediction coding \cite{c1} (LPC)-based  data augmentation strategy, LPC Augment, which perturbs the frequency spectrum peak locations of the speech signal. Inspired by the LPC-based speaker normalization framework described in \cite{a4}, we create a new augmentation method to simulate formant shifts due to dialectal differences in children's speech.  A key difference between the proposed method and past methods is its ability to shift formants independently of each other by applying a different random perturbation to each pole of the LPC synthesis filter.  While other data augmentation techniques like VTLP also seek to shift formant frequencies, they can currently only scale formants frequencies by pre-determined amounts.  This is insufficient in modeling several of the dialect-dependent formant shifts that appear in language.  For example, in Southern AAE, the vowel \textbackslash EY\textbackslash{}  has a lower first formant and higher second formant than in Standard American English \cite{b22}.  We evaluate the effectiveness of the proposed method, in comparison with other data augmentation techniques, in improving the performance of both a hidden Markov Model deep neural network (HMM-DNN)-based system and an end-to-end system for cross-dialect children's speech recognition.

\section{Datasets and Methods}
\subsection{Novel Datasets}

This paper introduces two novel children's speech datasets \footnote{http://www.seas.ucla.edu/spapl/projects/Jibo.html} that can be used for low-resource children's ASR tasks.
The first is the UCLA JIBO Kids' Dataset.
This dataset contains recordings of approximately 130 children between the ages of 4 and 7 years old, the critical age range for early acquisition of literacy.  The children were recorded while they performed educational exercises in reading and pronunciation (eg. picture-naming tasks). Each child was recorded in 3 sessions each lasting about 15 minutes. The children conversed with the social robot, Jibo \footnote{``Jibo Robot - He can't wait to meet you," Boston, MA, 2017. [Online]. Available: https://www.jibo.com}, following a protocol created by experts in early childhood education \cite{b26}. A facilitator was also present at each session and intervened verbally if the child had difficulty interacting with the social robot. Each child sat approximately two feet away from the robot with a microphone placed equidistantly between them. The children then were administered a portion of the Goldman Fristoe Test of Articulation-3 (GFTA3) \cite{b27}, a common oral assessment used by speech-language pathologists, as well as exercises in counting and spelling. All children recruited to the study lived in Southern California and were proficient in English. Many of these children spoke second languages at home. The audio was recorded by a Logitech C920 Webcam microphone with a sampling rate of 48kHz. 

The second dataset is the GSU Kids' Dataset.
This dataset contains recordings of approximately 160 children between the ages of 8 and 10 years old.  The children were recorded while performing educational exercises in reading, language, and pronunciation with a facilitator.  The children then were administered a portion of the GFTA, as well as other assessments used by speech-language pathologists, and exercises in counting and spelling. Each child was recorded in 5 sessions each lasting about 2 to 10 minutes.  All children recruited to the study lived in the Atlanta Georgia Area and were native English speakers.  The audio was recorded by a computer microphone with a sampling rate of 44.1kHz.

In both datasets, the recordings took place in offices at the students' schools during the school day and include background noise as one would find in a real use case.  In this paper, we use only the single word utterances in both datasets from the GFTA assessment in order to focus on the performance of the acoustic model alone (no language model used).  Approximately 5 hours of audio data from each dataset were used, and all speech samples were downsampled to 16kHz for the ASR experiments.

\subsection{The Proposed Data Augmentation Method}
\label{sec:format}
In this section, we propose a novel data augmentation technique. We use the notation from the linear prediction equation $s[n]-\sum_{k=1}^{P} a_ks[n-k] =e[n]$ where $s[n]$ is the windowed frame of the signal and $e[n]$ is the residual,  the prediction order, P, is estimated as $P=2F_{max}$ (in kHz) $+ 2$ where $F_{max}$ is one half the sampling frequency.

The algorithm is then given as follows assuming an all pole model. For each frame:
\begin{algorithmic}[1]
\State Compute the LPC coefficients, $a_1,a_2,...a_P$ of the windowed signal.
\State Compute the residual $e[n]$ as the result of passing $s[n]$ through the filter $A(z) = 1 - \sum_{k=1}^{P} a_k z^{-k}$
\State Solve for the complex conjugate roots, $r_k$, of the prediction filter polynomial $A(z)$

\State Compute the magnitude and phase of each root $r_k$
\State Multiply the phase of each complex conjugate pair of roots by a warping factor $w_k \forall k=1,2,...P$ where the warping factor values are chosen from a random uniform distribution $\in [x,y]$ once for each utterance and held constant across all frames of the utterance.
\State Recombine the magnitude with phase of each pole, creating the warped polynomial roots $\hat{r}_k = \lvert r_k \rvert * e^{j(w_k* \angle r_k)}$.  The warping does not affect the magnitude in order to ensure filter stability.
\State Determine the new prediction polynomial, $\hat{A}(z)$ as the polynomial whose roots are the warped prediction filter roots, $\hat{r}_k$
\State Create the perturbed output frame by passing the residual $e[n]$ through the filter with transfer function $1/\hat{A}(z)$

\end{algorithmic}

An example of the spectrum of signal perturbed with LPC Augment is shown in Figure 1.  A block diagram of the algorithm is shown in Figure 2.

\begin{figure}[h]
\centering
\includegraphics[width=0.5\textwidth, height=4cm]{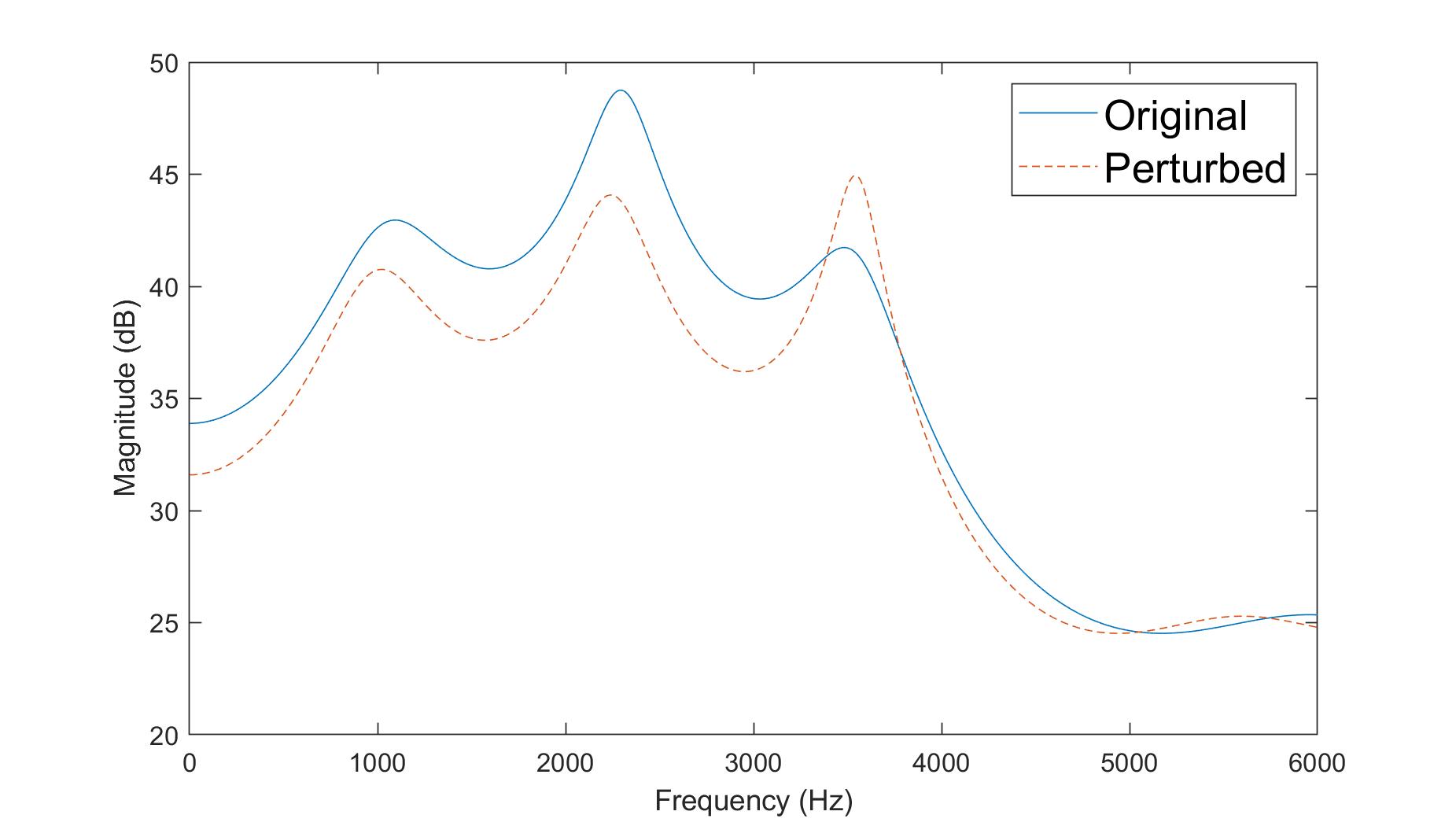}
\caption{An example of the LPC spectra of a child in the UCLA JIBO kid's Database pronouncing the phoneme \textbackslash AA\textbackslash{}, and the result of perturbing it with LPC Augment.  In the perturbed signal, the first two formant peaks have been shifted to the left, and the third has been shifted to the right.}
\end{figure}

\begin{figure}[ht]
\centering
\includegraphics[width=0.4\textwidth, height=12cm]{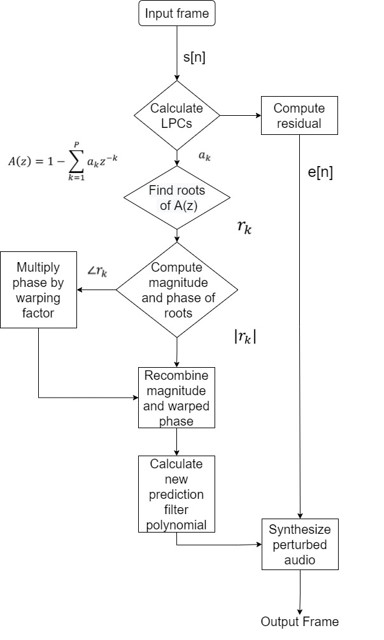}
\caption{Diagram of the LPC Augment Algorithm}
\end{figure}
\section{Experimental Setup}

In this section, we introduce the experimental setup including the data and models. For the data, both the Californa English speech data from the UCLA JIBO kids' dataset and Georgia English (we henceforth refer to the combination of Southern American English and African American English represented in the dataset as ``Georgia English") from the GSU kids' dataset are split into training, validation, and test sets with a ratio of 7:1:2 with no overlap between speakers. As a result, approximately 3.5 hours of data are used for training. We verify the proposed method across both a hybrid HMM-DNN model and an end-to-end attention-based model to demonstrate that LPC Augment is not model-dependent.

The hybrid system is built based on Kaldi \cite{b32} and Pykaldi2 \cite{b29}, where Pykaldi2 is used for acoustic model training. Specifically, a four-layer bidirectional long short term memory (BLSTM) model with 512 nodes in each direction is used. The input is 80 dimensional log filter-bank energy extracted with a frame length of 25ms and a frame shift of 10ms. We also use a frame skipping strategy \cite{b31} by concatenating two adjacent frames and then skipping the frame by a ratio of 2 to accelerate the model training. The acoustic model then outputs an approximately 3488 dimensional vector representing the senone probabilities for each frame, which are then decoded using a pre-constructed WFST graph in Kaldi. Prior to the acoustic model training for kids' data, HMM-GMM and BLSTM models are trained using the Librispeech clean 100 hours data \cite{b33}, as the model for the forced alignment acquisition and the start point for kids' acoustic model training, respectively. 

The end-to-end model is the sequence-to-sequence (S2S) Speech Transformer as proposed in \cite{b30}.  The model input is the spectrogram calculated with a frame size of 25ms and a frame shift of 10ms. The input is then passed to the network which consists of a series of three convolutional layers each with a receptive field of size 11, and an encoder and decoder block both composed of six stacked multi-head attention units and fully connected layers with residual connections. The output then contains 31 classes: 26 lowercase letters, apostrophe, period, space, noise marker, and end-of-sequence tokens.

We use the proposed data augmentation scheme to train the model and evaluate the performance across training and testing conditions. Data augmentation was performed using MATLAB's LPC coefficient algorithm.  Each utterance is windowed using a 20ms long Hamming window before being passed into the augmentation algorithm.  The length of the Hamming window was determined empirically in pilot experiments.  The perturbed audio sample is then input into the neural network.  We first optimize the range of the warping factors $w_k$ using the validation set. The training set size was increased by 3x with the proposed method.  Preliminary experiments showed that augmenting the training set size to 5x yielded no additional increase in performance. We then compare the performance of the optimized proposed method with that of other common data augmentation methods in recognizing children's speech of both the in-domain dialect and an out-of-domain dialect.  

We first train the models on speech from one of the English dialects and test on the other dialect for the zero resource scenario.  We then train a model jointly on both children's datasets for the low-resource scenario. 

\section{Results and Discussion}

\subsection{Optimizing the Warping Factor}

\begin{table}[htb]
\begin{center}
\begin{tabular}{||c | c | c||}

 \hline
  Warping Factor & CA val & GA val  \\ [0.5ex] 
 \hline
 $w \in [0.8, 1.0]$ & 16.41 & 63.44 \\ 
 \hline
 $ w \in [1.0, 1.2]$ & 15.97 & 69.16  \\
 \hline
$w \in [0.8, 1.2]$ & 14.63 & \textbf{51.63}  \\
 \hline
$ w \in [0.9, 1.1]$ & 14.86 & 55.93  \\
 \hline
$w \in [0.7, 1.3]$ & \textbf{13.94} & 58.85  \\ [1ex] 
 \hline
\end{tabular}
\caption{Results of the recognition experiment (in \%WER) on the validation set with the proposed method for different warping factors using the transformer model.  CA Val denotes the performance of the system trained with data augmentation on the speech data containing dialects found in Georgia and validated on speech containing dialects found in California.  GA Val similarly denotes the performance of the system trained with data augmentation on the speech data collected in California and validated on the speech data collected in Georgia. The lowest WER for each case is shown in boldface.}
\end{center}
\end{table}

\begin{table*}[t]
\begin{center}
\begin{tabular}{ ||c|c|>{\columncolor[gray]{0.8}}c|c|>{\columncolor[gray]{0.8}}c|>{\columncolor[gray]{0.8}}c|c|>{\columncolor[gray]{0.8}}c|c|| }
\hline
 & \multicolumn{4}{|c|}{Train CA} & \multicolumn{4}{|c||}{Train GA} \\
 \hline
 \%WER & \multicolumn{2}{|c|}{Transformer} & \multicolumn{2}{|c|}{HMM-DNN} & \multicolumn{2}{|c|}{Transformer} & \multicolumn{2}{|c||}{HMM-DNN} \\
 \hline
 
 Test Set & CA test & GA test & CA test & GA test & CA test & GA test & CA test & GA test  \\ [0.5ex] 
 \hline
 No Aug & 18.34 & 70.00 & 16.39 & 76.29 &  56.56 & 24.01 & 92.26 & 37.74\\ \hline VTLP \cite{b8} & 20.07 & 71.46
 & 15.85 & 77.83 & 65.41 & 25.10 & 91.26 & 37.29 \\
 
 \hline
 Speed Pert. \cite{b9} & 26.39 & 69.58  & 14.57 & 76.74 & 63.12 & 27.82 & 90.44 & 38.82 \\
 \hline
 SpecAug \cite{b11} & \textbf{17.85} & 62.84 & 13.93 & 76.47 &  54.84 & \textbf{22.64} & 88.71 & \textbf{34.84} \\
 \hline
 LPC Aug & 19.49 & 62.70 & 14.30 & 76.74 & 51.76 & 24.79 & 81.33 & 38.73\\
 \hline
 \makecell{Speed Pert. \\ + SpecAug} & 21.32 & 68.63 & 14.30 & 76.92 & 63.85 & 22.95 & 90.16 & 37.56  \\
 \hline
 \makecell{Speed Pert. \\ + LPC Aug} & 23.86 & 71.88 & 13.75 & 77.29  & 55.68 & 23.54 & 83.15 & 36.83 \\[1ex] 
 \hline
 \makecell{SpecAug \\ + LPC Aug} & 18.61 & \textbf{59.80} & \textbf{13.30} & \textbf{75.02}  & \textbf{51.13} &  22.90 & \textbf{75.41} & 35.48 \\ 
 \hline
\end{tabular}
\caption{Comparison of common speech data augmentation methods with the proposed method.  Each model (Transformer and HMM-DNN) is trained on either the California English training set (Train CA) or the Georgia English training set (Train GA) and then evaluated on both the California English test set (CA test) and the Georgia English test set (GA Test).  Columns representing zero-resource scenarios (where the model is trained on only one dialect and tested on the other) are highlighted.  The lowest word error rate for each case is shown in boldface.} 
\end{center}
\end{table*}

In order to optimize the range of warping factors, $w_k$, used in the proposed method, we first train the transformer model on the California English training set and evaluate it on both the California English and Georgia English validation set (CA val and GA val respectively).  We use the training set containing California English because it is considered a widely-spoken American dialect.  Adapting the California English training set to the Georgia English validation set then represents adapting from a more standard dialect to the less standard dialect as in low-resource scenarios.  Table 1 shows the performance in percent word error rate (\%WER) of the proposed algorithm for warping factors within the indicated range.

\subsection{Zero Resource Scenario}

Here, we are primarily concerned with achieving the best result on the out-of-domain data (GA Test), and so we continue with the warping factor chosen in the range $[0.8, 1.2]$.  Zero resource scenarios occur when the model is trained on only one dialect and tested on another. We proceed to compare the performance of the proposed method (abbreviated LPC Aug) in zero resource dialect children's ASR with three of the most commonly used data augmentation algorithms: VTLP, Speed Perturbation (Speed Pert.), and Spec Augment (SpecAug).  We also combine the more successful data augmentation methods to determine their cumulative effects. The results of both the transformer and hybrid model are shown in Table 2. The proposed method, LPC Augment, achieves a statistically significant ($p<0.05$) reduction in WER over the baseline (``No Aug") for mismatched dialect cases.  The lowest WER when training on one dialect and testing on the other is achieved when LPC Augment is used in conjunction with SpecAugment.  In testing and training on the same dialect, the lowest WER is achieved by using SpecAugment alone in three out of four cases.  

\subsection{Low Resource Scenario}

We train the models on data from both the CA dataset and the GA dataset in order to create an ASR system that performs well over multiple dialects and ages.  This represents the low-resource case, as the training sets from both dialects are small.  We show the results in \%WER in Table 3.  Note that the high baseline WERs observed in Tables 2 and 3 have been observed in previous low-resource accented children's ASR tasks in other languages as well \cite{b34}.

\begin{table}[htb]
\begin{center}
\begin{tabular}{ ||c|c|c|c|c|| }
 \hline
  & \multicolumn{2}{|c|}{Transformer} & \multicolumn{2}{|c||}{HMM-DNN} \\
 \hline
 
 \%WER & CA test & GA test & CA test & GA test  \\ [0.5ex] 
 \hline
 No Aug & 26.12
 & 21.18
 &  16.94 & 37.83\\ 
 \hline
  VTLP & 21.47
 & 15.84
 &  17.49 & 36.83\\ 
 \hline
 Speed Pert. & 19.68
 & 14.57
 & 15.76 & 37.92  \\
 \hline
 SpecAug & 19.23 & 13.80 & 15.03 & \textbf{35.20}  \\
  \hline
 LPC Aug & 19.76 & 14.39 & 14.66 & 38.46 \\
 \hline
 \makecell{Speed Pert. \\ + SpecAug} & \textbf{18.72} & 13.76 & 14.39 & 36.74  \\
 \hline
 \makecell{Speed Pert. \\ + LPC Aug} & 19.01 & 13.52 & 14.30 & 37.47  \\[1ex]
 \hline
 \makecell{SpecAug \\ + LPC Aug} & 18.91 & \textbf{13.34} & \textbf{13.84} & 35.29  \\
 \hline
 
\end{tabular}
\caption{Results of the models trained on both Train CA and Train GA and tested on CA Test and GA Test with the proposed and other data augmentation methods.  The lowest WER for each case is shown in boldface.}
\end{center}
\end{table}

\subsection{Discussion}

We observe that LPC Augment creates a significant reduction in WER for zero resource dialect children's ASR as compared to the other frequency-based data augmentation method, VTLP. This is likely due to the algorithm changing formant locations independently of each other rather than according to a predefined warping function. It appears that LPC Augment is complementary to SpecAugment, as they can typically be used together to give better performance than either alone or compared to other data augmentation methods.  In the zero-resource scenario in Table 2, the HMM-DNN ASR system results in a much higher reduction in WER for the CA test set than for the GA test set.  This may be a result of pre-training the model on Librispeech 100-clean which contains speech of a dialect more similar to California English.  Further work is necessary to determine how the pre-training dataset biases the model towards better performance for a given dialect.  We also notice in Table 3 (low-resource case) that the transformer model typically benefits more (\% improvement over the baseline) from data augmentation than the HMM-DNN system.  The transformer's implicit language modeling may allow it to better learn relevant groupings of characters and hence may have a bigger advantage for the children's small vocabulary task.  In the low-resource task, we observe that LPC Augment used simultaneously with SpecAugment and Speed Perturbation appears to give improved performance across dialects.  We conclude that LPC Augment shows promise in creating robust low and zero-resource dialect ASR systems.

\section{Conclusions}
The proposed LPC-based data augmentation scheme provides significant reduction in WER for children's out-of-domain dialect ASR.  The method can also be used in combination with SpecAugment to further improve ASR performance for both HMM-DNN and Transformer models.  We showed comparable or better performance than the state-of-the-art data augmentation methods for the zero resource case and were able to create a reliable system in the low-resource case.  All improvements were statistically significant.  Future work includes evaluating the algorithm across children of different ages and further optimizing the choice of warping factors for different age groups and dialects.  Future work also include evaluating the proposed method for adults' multi-dialect and low-resource ASR systems.

\section{Acknowledgements}

This work was supported in part by the NSF and the Ralph J. Bunche Center for African American Studies at UCLA.  The GSU data was collected with support by the Eunice Kennedy Shriver National Institute of Child Health \& Human Development of the NIH under Grant P01HD070837.


\label{sec:pagestyle}

\bibliographystyle{IEEEtran}

\bibliography{mybib}

\bibliographystyle{IEEEbib}
\bibliography{strings,refs}

\begin{thebibliography}{9}



\bibitem{s1} A. Koenecke et. al., ``Racial disparities in automated speech recognition," Proceedings of the National Academy of Sciences, 2020, 117 (14) pp. 7684-7689; DOI: 10.1073/pnas.1915768117

\bibitem{a1} H. Hu et al., ``REDAT: Accent-Invariant Representation for End-To-End ASR by Domain Adversarial Training with Relabeling," in \textit{ICASSP} 2021, pp. 6408-6412, doi: 10.1109/ICASSP39728.2021.9414291.
\bibitem{a2} J. Luo et al., ``Cross-Language Transfer Learning and Domain Adaptation for End-to-End Automatic Speech Recognition," in \textit{ICME}, 2021, pp. 1-6, doi: 10.1109/ICME51207.2021.9428334.
\bibitem{a3} T. Tan, Y. Lu, R. Ma, S. Zhu, J. Guo, and Y. Qian, "AISpeech-SJTU ASR System for the Accented English Speech Recognition Challenge," in \textit{ICASSP}  2021, pp. 6413-6417, doi: 10.1109/ICASSP39728.2021.9414471.

\bibitem{b8} N. Jaitly and G. E. Hinton, ``Vocal tract length perturbation (VTLP) improves speech recognition," in \textit{ICML Workshop on Deep Learning for Audio, Speech, and Language Processing,} 2013.
\bibitem{b9} T. Ko, V. Peddinti, D. Povey, and S. Khudanpur, ``Audio Augmentation for Speech Recognition," in \textit{INTERSPEECH,} 2015.
\bibitem{b11} D. S. Park, et al. ``SpecAugment: A Simple Data Augmentation
Method for Automatic Speech Recognition." in \textit{Proc. of Interspeech,}
2019, pp. 2613-2617.
\bibitem{b12} A. Hagen, B. Pellom, R. Cole, ``Highly accurate children's speech recognition for interactive reading tutors using subword units,” Speech Communication, Volume 49, Issue 12, 2007, pp. 861-873,
\bibitem{b13} L. L. Koenig, J. C. Lucero, and E. Perlman, ``Speech Production
Variability in Fricatives of Children and Adults: Results of Functional Data Analysis,” \textit{JASA}, vol. 124, no. 5, pp. 3158–3170, 2008.
\bibitem{b14} S. Lee, A. Potamianos, and S. Narayanan, ``Acoustics of Children’s Speech: Developmental Changes of Temporal and Spectral
Parameters,” The Journal of the Acoustical Society of America,
vol. 105, no. 3, pp. 1455–1468, 1999.
\bibitem{b15} S. Das, D. Nix, and M. Picheny, ``Improvements in Children’s
Speech Recognition Performance,” in \textit{Proc. of ICASSP,} 1998, pp. 433–436.
\bibitem{b16} G. Yeung and A. Alwan, ``On the Difficulties of Automatic Speech Recoginition for Kindergarten-Aged Children," in \textit{Interspeech} 2018, pp. 1661-1665.
\bibitem{b17} K. Shobaki, J.-P. Hosom, and R. A. Cole, ``The OGI kids' speech corpus and recognizers," in \textit{ICSLP,} 2000.
\bibitem{b19} J. Fainberg, P. Bell, M. Lincoln, and S. Renals, ``Improving Children’s Speech Recognition through Out-of-Domain Data Augmentation," in \textit{Proc. of INTERSPEECH,} 2016, pp. 1598–1602.
\bibitem{b20} J. Wang, Y. Zhu, R. Fan, W. Chu, and A. Alwan, ``low-resource German ASR with Untranscribed Data Spoken by Non-native Children – INTERSPEECH 2021 Shared Task SPAPL System," in \textit{Proc. of Interspeech 2021}, pp. 1279-1283, doi: 10.21437/Interspeech.2021-1974.
\bibitem{c1} J. Haskew, J. Kelly, R. Kelly and T. McKinney, ``Results of a Study of the Linear Prediction Vocoder," in \textit{IEEE Transactions on Communications}, vol. 21, no. 9, pp. 1008-1015, September 1973, doi: 10.1109/TCOM.1973.1091782.
\bibitem{a4} H. Kumar Kathania, S. Reddy Kadiri, P. Alku and M. Kurimo, ``Study of Formant Modification for Children ASR," in \textit{ICASSP} 2020, pp. 7429-7433, doi: 10.1109/ICASSP40776.2020.9053334.
\bibitem{b22} W. A. Kretzschmar, Jr., ``African American Voices in Atlanta," in \textit{The Oxford handbook of African American language,} ed. by Jennifer Bloomquist, Lisa J. Green, and Sonja L. Lanehart. Oxford: Oxford University Press.  2015.


\bibitem{b26} G. Yeung et al., ``A robotic interface for the administration of language, literacy, and speech pathology assessments for children", SLATE, 2019, pp. 41-42. 

\bibitem{b27} R. Goldman and M. Fristoe, ``Gfta 3: Goldman Fristoe 3 Test of Articulation," 2015.

\bibitem{b32} D. Povey et al., ``The Kaldi Speech Recognition Toolkit," \textit{in: IEEE
ASRU}, 2011


\bibitem{b29} L. Lu, X. Xiao, Z. Chen, and Y. Gong, ``PyKaldi2:
Yet another speech toolkit based on Kaldi and PyTorch,"
arXiv:1907.05955, 2019.

\bibitem{b31} Y. Miao, J. Li, Y. Wang, S. Zhang, and Y. Gong, ``Simplifying long short-term memory acoustic models for fast training and de- coding,” in \textit{Proc. ICASSP,} 2016.

\bibitem{b33} V. Panayotov, G. Chen, D. Povey, S. Khudanpur, 2015. ``Librispeech: An ASR corpus based on public domain audio books", \textit{in: Proc. IEEE ICASSP}, 2015, pp. 5206–5210. doi:10.1109/ICASSP.2015.7178964

\bibitem{b30} L. Dong, S. Xu and B. Xu, ``Speech-Transformer: A No-Recurrence Sequence-to-Sequence Model for Speech Recognition," in \textit{ICASSP,} 2018, pp. 5884-5888, doi: 10.1109/ICASSP.2018.8462506.

\bibitem{b34} Gretter, R., Matassoni, M., Falavigna, D., Misra, A., Leong, C.W., Knill, K., Wang, L. (2021) ``ETLT 2021: Shared Task on Automatic Speech Recognition for Non-Native Children’s Speech". Proc. Interspeech 2021, 3845-3849, doi: 10.21437/Interspeech.2021-1237









\end{thebibliography}

\end{document}